\documentclass{article}

\usepackage{arxiv}

\usepackage[utf8]{inputenc} 
\usepackage[T1]{fontenc}    

\usepackage[numbers,sort&compress]{natbib}

\usepackage{url}            
\usepackage{booktabs}       
\usepackage{amsfonts}
\usepackage{amsmath,amssymb}
\usepackage{nicefrac}       
\usepackage{microtype}      
\usepackage{graphicx}
\usepackage{doi}
\usepackage{tikz}
\usetikzlibrary{arrows.meta,positioning,fit,shapes.geometric}
\usepackage[caption=false,font=footnotesize]{subfig}
\usepackage{xcolor}               
\usepackage{tabularx}
\usepackage{array}
\usepackage{placeins}      

\usepackage{algorithm}      
\usepackage{algpseudocode}  

\usepackage{hyperref}       
\usepackage{cleveref}       

\newcolumntype{Y}{>{\raggedright\arraybackslash}X}

\title{Topology-Aware Hybrid Wi-Fi/BLE Fingerprinting via Evidence-Theoretic Fusion and Persistent Homology}

\newif\ifuniqueAffiliation
\uniqueAffiliationtrue   

\ifuniqueAffiliation 
\author{Behrad~Mousaei~Shir-Mohammad, Behzad~Moshiri and Abolfazl~Yaghmaei
	\thanks{Manuscript received XXXX 2025; revised XXXX 2025.
		This work was supported in part by the U.S. Department of Commerce under Grant BS123456.}%
	\thanks{The authors are with the School of Electrical and Computer Engineering, University of Tehran, Iran, Tehran 1439957131
		(e-mail: \texttt{\{behrad.mousaie,yaghmaei,moshiri\}@ut.ac.ir}).}
	\thanks{All code and the dataset for this paper are available at \url{https://github.com/behradbx/Tehran-Indoor-Localization}.}
}
\else
\usepackage{authblk}

\newbox{\orcid}\sbox{\orcid}{\includegraphics[scale=0.06]{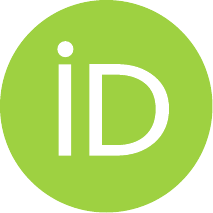}}
\author[1]{%
	\href{https://orcid.org/0000-0000-0000-0000}{\usebox{\orcid}\hspace{1mm}David S.~Hippocampus\thanks{\texttt{hippo@cs.cranberry-lemon.edu}}}%
}
\author[1,2]{%
	\href{https://orcid.org/0000-0000-0000-0000}{\usebox{\orcid}\hspace{1mm}Elias D.~Striatum\thanks{\texttt{stariate@ee.mount-sheikh.edu}}}%
}
\affil[1]{Department of Computer Science, Cranberry-Lemon University, Pittsburgh, PA 15213}
\affil[2]{Department of Electrical Engineering, Mount-Sheikh University, Santa Narimana, Levand}
\fi

\renewcommand{\shorttitle}{\textit{arXiv} Template}

\hypersetup{
	pdftitle={Topology-Aware Hybrid Wi-Fi/BLE Fingerprinting via Evidence-Theoretic Fusion and Persistent Homology},
	pdfsubject={Indoor localization, Wi-Fi/BLE, data fusion},
	pdfauthor={Behrad Mousaei Shir-Mohammad, Behzad Moshiri, Abolfazl Yaghmaei},
	pdfkeywords={Indoor localization, Wi-Fi fingerprinting, BLE beacons, Data Fusion, Discrete calculus, Kalman filter, particle filter, sensor fusion, Dempster--Shafer theory, Choquet integral, persistent homology, uncertainty quantification, embedded systems},
	colorlinks=true,
	linkcolor=blue,
	citecolor=blue,
	urlcolor=blue
}

\begin{document}
	\maketitle
	
	\begin{abstract}
		Indoor localization remains challenging in GNSS-denied environments due to multipath, device heterogeneity, and volatile radio conditions. We propose a topology-aware, hybrid Wi-Fi/BLE fingerprinting framework that (i) applies physically consistent RSS normalization (dBm $z$-scoring or dBm$\!\to$linear mW$\!\to z$-score), (ii) denoises streams with classical Bayesian filters (KF/UKF/PF), (iii) combines complementary regressors (Random Forest and weighted kNN with a diagonal Mahalanobis metric), (iv) performs evidence-theoretic fusion via Dempster–Shafer theory (DST), and (v) augments each sample with persistent-homology (PH) descriptors. The system outputs both $(x,y)$ estimates and interpretable belief maps, and is engineered for microcontroller-class deployment with per-update cost $O(T\log M+\log M+M_p+S)$.
		
		We evaluate on two heterogeneous datasets—including a new 1{,}200-sample ESP32 survey—and report ablations, robustness to test-only noise, and significance across 10 stratified splits. Under \emph{10\% synthetic RSS noise}, the full pipeline attains \textbf{3.40\,m} (Dataset~1) and \textbf{2.45\,m} (Dataset~2) RMSE, improving a strong PF\,+\,RF baseline by about \textbf{37\%}. Averaged across splits, it yields \textbf{$4.993\pm0.15$\,m} versus \textbf{$6.292\pm0.13$\,m} (20.6\% relative reduction; $p<0.001$). In \emph{noise-free} tests, accuracy tightens to \textbf{0.44\,m} and \textbf{0.32\,m} (up to 56\% better). Compared with recent learning-heavy approaches that assume large site-specific datasets and GPU inference, our method delivers competitive accuracy with formal uncertainty quantification and low computational cost suitable for real-time deployment.
	\end{abstract}
	
	\keywords{
		Indoor localization, Wi-Fi fingerprinting, BLE beacons, Data Fusion, Discrete calculus, Kalman filter, particle filter, sensor fusion, Dempster–Shafer theory, Choquet integral, persistent homology, uncertainty quantification, embedded systems.
	}
	
	
	\section{Introduction}
	\label{sec:introduction}
	
	Global Navigation Satellite Systems (GNSS) degrade or fail indoors, yet most buildings already host dense Wi-Fi access points and, increasingly, Bluetooth Low Energy (BLE) beacons. RSS fingerprinting leverages these ambient radios by matching received-signal-strength (RSS) vectors to a surveyed database of reference points (RPs) \cite{Zafari2019,Sesyuk2022}. In practice, however, multipath, device heterogeneity, and transient interference inject substantial noise into RSS, which destabilises distance metrics and degrades position estimates. Our goal is a lightweight, interpretable pipeline that explicitly addresses these realities while remaining deployable on commodity microcontrollers.
	
	\begin{figure*}[!t]
		\centering
		\resizebox{\textwidth}{!}{%
			\begin{tikzpicture}[
				node distance=5mm and 8mm,
				box/.style={rectangle, draw, rounded corners, align=left, inner sep=3pt, very thick},
				smallbox/.style={rectangle, draw, rounded corners, align=left, inner sep=2.2pt, thick, font=\scriptsize},
				arr/.style={-{Latex[length=2.6mm]}, very thick},
				lab/.style={font=\scriptsize, inner sep=0.5pt},
				dashedbox/.style={rectangle, draw, dashed, rounded corners, inner sep=5pt, thick},
				every node/.style={font=\footnotesize}
				]
				
				\node[box, text width=0.15\textwidth] (acq) {%
					RSS acquisition\\
					Wi-Fi (7 APs), BLE (3 beacons)\\
					Raw $\mathbf f\in\mathbb R^{d}$ per scan
				};
				
				\node[box, right=of acq, text width=0.19\textwidth] (norm) {%
					Normalisation\\
					$\tilde{\mathbf f}=(\mathbf f-\boldsymbol\mu)\oslash\boldsymbol\sigma$\\
					or $\mathbf p=10^{\mathbf f/10}$ mW,\;
					$\tilde{\mathbf f}=(\mathbf p-\boldsymbol\mu_p)\oslash\boldsymbol\sigma_p$
				};
				
				\node[box, right=of norm, text width=0.20\textwidth] (filter) {%
					RSS denoising (KF/UKF/PF)\\
					KF:\; $\mathbf x_t=\mathbf x_{t-1}+\mathbf v_{t-1}$,\;
					$\tilde{\mathbf z}_t=\mathbf x_t+\mathbf n_t$\\
					PF:\; $w_t^{(m)}\!\propto\!p(\tilde{\mathbf z}_t\!\mid\!\mathbf x_t^{(m)})$;\;
					resample if $M_{\mathrm{eff}}<0.3M_p$
				};
				
				\node[box, right=of filter, text width=0.22\textwidth] (feat) {%
					Feature assembly\\
					Fingerprint $\tilde{\mathbf z}\in\mathbb R^{d}$\\
					Per-sample PH on $\{(i,\tilde f_i)\}_{i=1}^{d}$ \\
					$\phi_{\mathrm{PH}}=[\mathrm{NoP}_0,\mathrm{PE}_0,\mathrm{NoP}_1,\mathrm{PE}_1]$ with
					$\mathrm{PE}_k=-\sum_i \tilde\ell_i\log\tilde\ell_i$,
					$\tilde\ell_i=\dfrac{d_i-b_i}{\sum_j(d_j-b_j)}$\\
					Augment:\; $\mathbf x=[\tilde{\mathbf z};\,\phi_{\mathrm{PH}}]\in\mathbb R^{d+4}$
				};
				
				\node[smallbox, right=10mm of feat, yshift=6mm, text width=0.16\textwidth] (rf) {%
					Random Forest (RF)\\
					$\hat{\mathbf y}_{\mathrm{RF}}=\dfrac{1}{T}\sum_{t=1}^{T} h_t(\mathbf x)$
				};
				
				\node[smallbox, right=10mm of feat, yshift=-14mm, text width=0.16\textwidth] (knn) {%
					Weighted $k$NN (wKNN)\\
					$\delta(\mathbf x,\mathbf f_r)=(\mathbf x-\mathbf f_r)^{\!\top}\boldsymbol\Sigma^{-1}(\mathbf x-\mathbf f_r)$\\
					$w_r\propto 1/(\delta+\epsilon)$,\;
					$\hat{\mathbf y}_{\mathrm{KNN}}=\dfrac{\sum_r w_r\,\mathbf y_r}{\sum_r w_r}$
				};
				
				\node[smallbox, right=14mm of rf, yshift=-4mm, text width=0.22\textwidth] (dst) {%
					Dempster--Shafer (DST) fusion\\
					Cells $\Theta=\{\theta_j\}_{j=1}^{S}$,\; $m_s(\{\theta_j\})\propto e^{-\alpha\,\delta_s(j)}$\\
					$m(A)=\dfrac{\sum_{B\cap C=A} m_1(B)m_2(C)}{1-\sum_{B\cap C=\varnothing} m_1(B)m_2(C)}$
				};
				
				\node[box, right=12mm of dst, text width=0.18\textwidth] (belief) {%
					Belief map \& estimate\\
					Belief $m(\cdot)$ over cells;\; $\theta^\star=\arg\max_{A\subseteq\Theta} m(A)$\\
					Point $\hat{\mathbf y}$ from cell centroid or fused regression\\
					Uncertainty: highest-belief regions
				};
				
				\draw[arr] (acq) -- (norm);
				\draw[arr] (norm) -- (filter);
				\draw[arr] (filter) -- (feat);
				
				\draw[arr] (feat) -- ([xshift=-2mm]rf.west);
				\draw[arr] (feat) -- ([xshift=-2mm]knn.west);
				
				\draw[arr] (rf.east) -- (dst.west);
				\draw[arr] (knn.east) -- (dst.west);
				
				\draw[arr] (dst.east) -- (belief.west);
				
				\node[dashedbox, fit=(norm)(feat), label={[lab]above:Offline (normalisation stats, $\boldsymbol\Sigma$, RF training)}] (offline) {};
				\node[dashedbox, fit=(acq)(filter)(rf)(knn)(dst)(belief), label={[lab]below:Online (per-scan; cost $\mathcal{O}(T\log M+\log M+M_p+S)$)}] (online) {};
				
			\end{tikzpicture}%
		} 
		\caption{End-to-end hybrid Wi-Fi/BLE fingerprinting pipeline. RSS is normalised, optionally denoised (KF/UKF/PF), and augmented with per-sample persistent-homology descriptors. Two complementary regressors (RF and wKNN) produce predictions that are fused by Dempster--Shafer (DST), yielding a point estimate and an interpretable belief map. Offline: learn normalisation statistics, $\boldsymbol\Sigma$, and RF. Online: per-scan operation with overall cost $\mathcal{O}(T\log M+\log M+M_p+S)$.}
		\label{fig:pipeline}
	\end{figure*}
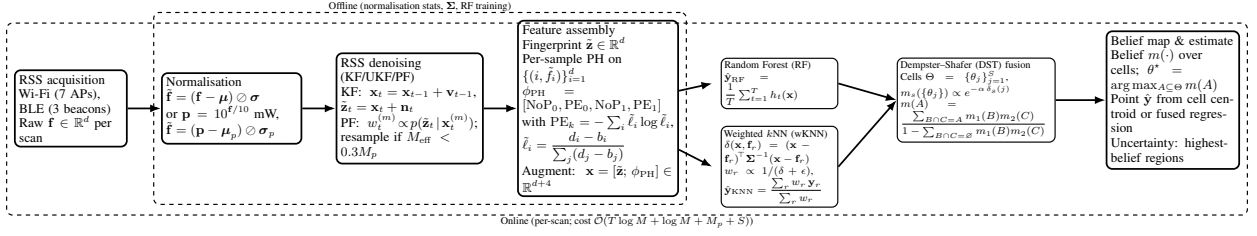
	
	In the broader context of sensor fusion and state estimation, prior work has demonstrated the effectiveness of Kalman-family filters and nonlinear blending for noisy, heterogeneous sensors: GPS/INS integration with nonlinear blending filters \cite{Rezaie2007}, MEMS-IMU calibration with Kalman filtering \cite{Jafari2015}, and robust fault handling in microgrids using the Unscented Kalman filter \cite{Vafamand2021}. These results motivate our choice of KF/UKF/PF to stabilise RSS before regression, providing accuracy gains with modest computational cost.
	
	Prior indoor-localisation work spans three strands. (i) \emph{Lightweight Wi-Fi–only methods} (e.g., selective/range-limited $k$NN) are easy to deploy but typically plateau at several metres and rarely quantify uncertainty; for example, SLWKNN reports $\sim$4.8\,m and DumbLoc $\sim$6.2\,m in noise-free conditions, without explicit RSS noise modelling or multi-radio fusion \cite{Leng2024,Narasimman2024}. (ii) \emph{Classical learning methods} improve regressors or features (e.g., RF-assisted AP selection with kernel ELM) and achieve roughly 3.9\,m, but still rely on a single radio and lack uncertainty handling \cite{Hou2024}. (iii) \emph{Learning-heavy models} learn complex radio–environment mappings and can reach 1–2\,m using CNN–LSTM or high-order GNNs \cite{Wang2020DLoc,Kang2023HoGNN}; extensions with data augmentation and temporal modelling further push accuracy, e.g., CSI trajectory fingerprinting with GAN-augmented CNN–LSTM \cite{Zhang2021CSI} and generative hidden-Markov Wi-Fi fingerprinting attaining sub-meter accuracy under controlled conditions \cite{Belmonte2021Generative}. In parallel, Xia \emph{et~al.} optimise RSSI fingerprint matching in wireless sensor networks by refining the matching scheme for efficiency and accuracy \cite{Xia2025}. While effective, these approaches are \emph{single-modality} and typically do not expose uncertainty or perform evidence-theoretic fusion; by contrast, our hybrid Wi-Fi+BLE design integrates DST/Choquet fusion and PH features, and targets microcontroller-class deployment. Hybrid sensing has also been explored: SPOTTER fuses Wi-Fi and BLE in a particle-filter framework and reports $\sim$2.7\,m, but with comparatively heavy computation and without demonstration on resource-constrained hardware \cite{Azaddel2023}. Complementary studies of multi-sensor fusion show that Dempster--Shafer evidence theory is effective under conflicting/uncertain measurements, and fuzzy aggregation can capture non-linear complementarity \cite{HoseinNezhad2002,Kazemi2021}. These gaps motivate a solution that pairs multi-modal sensing and principled fusion with efficient, classical components.
	
	We therefore propose a lightweight, real-time hybrid Wi-Fi/BLE fingerprinting framework that integrates four components: (i) RSS filtering via Kalman, Unscented Kalman, or Particle filters to attenuate fast fading \cite{Konatowski2016}; (ii) dual regressors—Random Forest (RF) and weighted $k$-nearest neighbours (wKNN)—to produce complementary coordinate predictions; (iii) model fusion using Dempster--Shafer Theory (DST) for principled evidence combination and a two-source Choquet fuzzy integral to capture non-linear complementarity \cite{Sentz2002,Murofushi2000}; and (iv) topology-aware features computed \emph{per sample} via persistent homology (PH) to summarise channel-structure geometry beyond variance-based statistics \cite{Carlsson2009}. DST yields spatial belief maps visualised as highest-belief regions rather than Gaussian ellipses, improving operator interpretability. RF and wKNN respond differently to noise and RP sparsity, so fusing them stabilises predictions across sites; filtering reduces measurement noise at the source; and PH descriptors capture coarse geometric structure that PCA-like features may miss. All components are commodity, require no GPUs, and suit embedded deployment. Our design is informed by successes of nonlinear Kalman filtering and fuzzy aggregation in related sensing domains \cite{Rezaie2007,Jafari2015,Vafamand2021,HoseinNezhad2002,Kazemi2021}.
	
	\subsection*{Problem Statement and Notation}
	\label{subsec:problem-notation}
	We consider indoor fingerprinting on a bounded floor plan. At time $t$, the sensor acquires a Wi-Fi/BLE RSS vector $\mathbf f_t\in\mathbb{R}^{d}$ across $N_W$ Wi-Fi and $N_B$ BLE channels ($d=N_W+N_B$). Let $\mathcal{D}_{\text{train}}=\{(\mathbf f^{(n)},\mathbf r^{(n)})\}_{n=1}^M$ be the labelled fingerprint–location pairs at surveyed reference points (RPs), with $\mathbf r^{(n)}=(x^{(n)},y^{(n)})\in\mathbb{R}^2$. The goal is to estimate $\hat{\mathbf r}_t=(\hat x_t,\hat y_t)$ for a new $\mathbf f_t$, while also exposing spatial uncertainty via a belief map over discretised cells $\Theta=\{\theta_j\}_{j=1}^S$.
	
	We normalise fingerprints either in dBm (direct $z$-score) or via dBm$\to$linear power (mW)$\to z$-score (Sec.~\ref{subsec:data}). For denoising, we apply KF/UKF/PF to the \emph{per-feature} temporal stream (Sec.~\ref{subsec:filter}). Two complementary regressors—RF and variance-weighted wKNN—produce coordinate predictions (Sec.~\ref{subsec:regression}), which are fused by Dempster--Shafer theory (DST) on $\Theta$; a two-source Choquet integral aggregates confidences (Sec.~\ref{subsec:fusion}). We optionally augment each input with four persistent-homology (PH) descriptors computed on the 1-D embedded curve of the normalised fingerprint (Sec.~\ref{subsec:tda}).
	
	\paragraph*{Contributions}
	\begin{itemize}
		\item A unified, compute-efficient Wi-Fi/BLE fingerprinting pipeline combining classical filtering, complementary regressors, evidence-theoretic fusion, and per-sample PH features—producing both point estimates and interpretable belief maps \cite{Sentz2002,Carlsson2009}.
		\item A physically meaningful feature normalisation: either direct $z$-scoring in dBm or dBm$\rightarrow$linear power (mW)$\rightarrow z$-score; we avoid $\ln|\cdot|$ on dBm, which lacks physical meaning.
		\item A transparent per-update complexity of $O(T\log M+\log M+M_p+S)$ for $T$ trees (RF), $M$ training fingerprints, $M_p$ PF particles (if enabled), and $S$ spatial cells for DST—consistent with real-time use on microcontroller-class processors.
		\item A comprehensive evaluation on two heterogeneous datasets—including a new 1{,}200-sample ESP32 survey—and ablations isolating each module under both noise-free and 10\% synthetic-noise conditions \cite{Moradbeikie2024}.
	\end{itemize}
	
	\paragraph*{Results in brief}
	Under the unified 10\% noise definition (Sec.~\ref{subsec:protocol}), the full pipeline attains 3.40\,m (Dataset~1) and 2.45\,m (Dataset~2) RMSE on the canonical held-out split, improving over a strong PF+RF baseline by about 37\%. In noise-free conditions, accuracy tightens to 0.44\,m and 0.32\,m, up to 56\% better than the same baseline. Averaged across 10 stratified splits with the same 10\% noise, the mean RMSE is $4.993\,\pm\,0.15$\,m versus $6.292\,\pm\,0.13$\,m for PF+RF (20.6\% reduction; Table~\ref{tab:ci_rmse}). This explains why Table~\ref{tab:rmse_noise} (canonical split) and Table~\ref{tab:ci_rmse} (10-split average) report different absolute values under the same noise setting.
	
	\paragraph*{Scope and fairness}
	Where many deep models assume site-specific large training sets and GPU-class compute \cite{Wang2020DLoc,Kang2023HoGNN,Zhang2021CSI,Belmonte2021Generative}, our focus is robust performance on commodity radios and embedded hardware. We report both noisy and noise-free results and clearly distinguish them (several comparative works publish only noise-free figures \cite{Leng2024,Narasimman2024,Hou2024}). Uncertainty is reported as belief maps via DST \cite{Sentz2002}.
	
	\paragraph*{Paper organization}
	Section~\ref{sec:methodology} presents the full pipeline—normalisation, RSS filtering, regression, fusion, and per-sample PH features—together with a brief complexity analysis. Section~\ref{sec:results} describes the datasets and protocol, reports ablations and statistical tests, and benchmarks against recent work. Section~\ref{sec:conclusion} summarises the findings and outlines future directions.

	
	\section{Proposed Methodology}
	\label{sec:methodology}
	
	The proposed hybrid Wi-Fi/BLE framework comprises an \emph{offline} fingerprint–map construction stage and an \emph{online} real-time positioning stage. Figure~\ref{fig:pipeline} gives a high-level view of the pipeline; details follow.
	
	\subsection{Data Collection and Pre-processing}
	\label{subsec:data}
	
\begin{figure}[t]
		\centering
		\includegraphics[width=\linewidth]{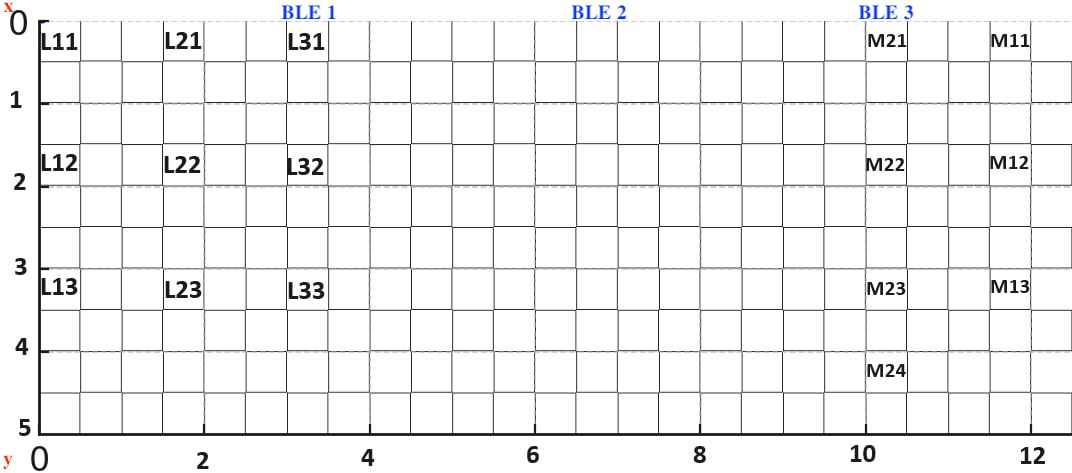}
		\caption{Map of Test area for Dataset~1 on the third floor of the Faculty of Electrical and Computer Engineering, University of Tehran. Fifteen reference points are \emph{distributed across} a $6\,\mathrm{m}\times14\,\mathrm{m}$ area ($50$\,cm each lattice).}
		\label{fig:floorplan}
	\end{figure}

	\textbf{Environments and datasets.}
	Two heterogeneous datasets are used. Dataset~1 was collected on the third floor of the Faculty of Electrical and Computer Engineering, University of Tehran (see Fig.~\ref{fig:floorplan}). We surveyed \textbf{15 reference points (RPs)} distributed over a \textbf{$6$\,m $\times$ $14$\,m} area (not a $50$\,cm lattice). At each RP we recorded RSS from \textbf{7 Wi-Fi APs} and \textbf{3 BLE beacons} using commodity ESP32 boards (one scanning Wi-Fi, one BLE). To reduce fast fading, we aggregate consecutive windows of \textbf{10 scans} into a single fingerprint; across the site this yields \textbf{$1{,}200$ fingerprints} (80 windowed fingerprints per RP). Dataset~2~\cite{Moradbeikie2024} contains \textbf{16 RPs} with \textbf{5 Wi-Fi} and \textbf{3 BLE} channels and \textbf{$1{,}400$ fingerprints}, built with the same windowing rule.
	
	\textbf{Fingerprint vector and normalisation.}
	Let $N_W$ and $N_B$ denote the numbers of Wi-Fi and BLE channels. The fingerprint at RP $r$ is
	\begin{equation}
		\mathbf f_r=\bigl[\rho^{(1)}_r,\dots,\rho^{(N_W)}_r,\;\beta^{(1)}_r,\dots,\beta^{(N_B)}_r\bigr]^\top\in\mathbb R^{d}
	\end{equation}
	\begin{equation}
		d=N_W+N_B.
	\end{equation}
	We avoid applying $\ln|\cdot|$ to dBm values (not physically meaningful). Instead, we use one of two physically consistent normalisations:
	\emph{(i)} direct $z$-scoring in dBm with statistics from the training set,
	\begin{equation}
		\tilde{\mathbf f}_r=(\mathbf f_r-\boldsymbol\mu)\oslash\boldsymbol\sigma ,
	\end{equation}
	or \emph{(ii)} conversion to linear power before $z$-scoring,
	\begin{equation}
		\mathbf p_r=10^{\mathbf f_r/10}\;(\mathrm{mW}),\qquad \tilde{\mathbf f}_r=(\mathbf p_r-\boldsymbol\mu_p)\oslash\boldsymbol\sigma_p .
	\end{equation}
	Unless stated otherwise we adopt the dBm $z$-score variant. The same statistics are applied to validation/test data.
	
	At each RP we collected raw scans at a fixed sampling cadence and averaged consecutive windows of 10 scans into a single fingerprint (80 fingerprints per RP on Dataset~1). The device height/pose and human presence were kept consistent across RPs to reduce occlusion bias; scans were performed in multiple short sessions to mitigate temporal drift.%
	All normalisation statistics ($\boldsymbol\mu,\boldsymbol\sigma$ or $\boldsymbol\mu_p,\boldsymbol\sigma_p$) are estimated \emph{on training data only} and reused for validation/test to prevent leakage.
	
	\subsection{Regression Models}
	\label{subsec:regression}
	
	We considered Linear Regression, SVR (RBF), Decision Tree, Random Forest (RF), and weighted $k$NN (wKNN). Hyper-parameters are tuned by five-fold cross-validation; the final ensemble retains \textbf{RF} and \textbf{wKNN} as complementary predictors that map $\tilde{\mathbf f}$ to 2-D coordinates $\hat{\mathbf r}=(\hat x,\hat y)$.
	
	\textbf{Distance metric for wKNN.}
	During prediction, we use a diagonal Mahalanobis-type distance in fingerprint space to equalise Wi-Fi/BLE dimensions:
	\begin{equation}
		\delta(\tilde{\mathbf z},\tilde{\mathbf f}_r)
		=(\tilde{\mathbf z}-\tilde{\mathbf f}_r)^\top\boldsymbol\Sigma^{-1}(\tilde{\mathbf z}-\tilde{\mathbf f}_r),
	\end{equation}
	where $\boldsymbol\Sigma=\mathrm{diag}(\hat\sigma_1^2,\ldots,\hat\sigma_d^2)$ are channel-wise variances estimated on the training set with $\ell_2$-shrinkage ($10^{-6}$ added to the diagonal) for numerical stability. wKNN uses inverse-distance weights; RF follows standard bagging over $T$ trees.
	
	Channel variances differ across Wi-Fi and BLE; scaling by $\boldsymbol\Sigma=\mathrm{diag}(\hat\sigma_1^2,\ldots,\hat\sigma_d^2)$ equalises dimensions and reduces dominance by a few volatile channels. We estimate $\hat\sigma_i^2$ on the \emph{training set} with $\ell_2$-shrinkage ($+10^{-6}$) to improve conditioning, then keep $\boldsymbol\Sigma$ fixed at test time.
	
	\subsection{RSS Noise Filtering}
	\label{subsec:filter}
	
	Filtering is applied \emph{independently per RSS channel} (scalar states), which is sufficient in our setting and reduces computational burden. For KF/UKF we set the measurement variance of channel $i$ to $R_i=\hat\sigma_i^2$ from the training set, and choose $Q_i=\gamma R_i$ with $\gamma\in\{0.25,0.5,1\}$ (Table~\ref{tab:hparams}). For PF, the likelihood uses the same $R_i$ and a unit-variance random walk in prediction; effective-sample-size (ESS) triggers systematic resampling (Alg.~\ref{alg:pf}).
	
	Instantaneous RSS vectors $\tilde{\mathbf z}_t$ are smoothed prior to regression using classical Bayesian filters~\cite{Konatowski2016}:
	
	\begin{enumerate}
		\item \textbf{Kalman Filter (KF):} linear–Gaussian state
		$\mathbf x_t=\mathbf x_{t-1}+\mathbf v_{t-1}$,\;
		$\tilde{\mathbf z}_t=\mathbf x_t+\mathbf n_t$,
		with $\mathbf v_{t-1}\!\sim\!\mathcal N(\mathbf 0,\mathbf Q)$ and
		$\mathbf n_t\!\sim\!\mathcal N(\mathbf 0,\mathbf R)$.
		\item \textbf{Unscented KF (UKF):} sigma-point prediction to handle mild non-linearities.
		\item \textbf{Particle Filter (PF):} non-parametric recursion with $M_p$ particles
		$\{(\mathbf x_t^{(m)},w_t^{(m)})\}_{m=1}^{M_p}$; resampling occurs when $\mathrm{ESS}=1/\sum_m (w_t^{(m)})^2<\tau\,M_p$.
	\end{enumerate}
	
	PF attained the lowest error in our experiments; KF provides a favourable accuracy/latency trade-off for embedded deployment.
	
	\subsection{Multi-Model Fusion and Uncertainty}
	\label{subsec:fusion}
	\paragraph*{DST grid and distance mapping.}
	We tile the floor with square cells of width $h\in\{0.5,0.75,1.0\}$\,m (Table~\ref{tab:hparams}). Each regressor $s$ outputs a point estimate $\hat{\mathbf r}_s$; we compute distances to cell centroids $\delta_s(j)=\|\hat{\mathbf r}_s-\mathbf c_j\|_2^2$ and convert them to singleton masses via Eq.~(5) with scale $\alpha$. Smaller $h$ increases $S$ and yields finer belief maps at a modest $O(S)$ runtime cost (Eq.~(12)).
	
	\paragraph*{Confidence scores for Choquet.}
	We define per-regressor confidences as $s_s=\max_j c_s(j)$ with $c_s(j)=\exp\{-\beta\delta_s(j)\}$, where $\beta$ is set to the inverse of the median validation-set distance (per dataset). This yields $s_s\in[0,1]$ that increases as the predicted point aligns tightly with a few high-belief cells.
	
	We fuse RF and wKNN predictions and quantify uncertainty using Dempster–Shafer Theory (DST)~\cite{Sentz2002} and a two-source Choquet fuzzy integral~\cite{Murofushi2000}.
	
	\paragraph*{Dempster–Shafer evidence fusion.}
	Discretise the environment into $S$ cells with centers $\{\mathbf c_j\}_{j=1}^S$ and frame of discernment $\Theta=\{\theta_1,\ldots,\theta_S\}$. Each regressor $s\in\{\mathrm{RF},\mathrm{KNN}\}$ yields a point estimate $\hat{\mathbf r}_s$ and induces a basic-belief assignment (BBA) $m_s:2^\Theta\!\to[0,1]$ by mapping distances to singleton masses:
	\begin{equation}
		\delta_s(j)=\|\hat{\mathbf r}_s-\mathbf c_j\|_2 ,\qquad
		m_s(\{\theta_j\})=\frac{\exp\!\bigl(-\alpha\,\delta_s(j)\bigr)}{\sum_{i=1}^S \exp\!\bigl(-\alpha\,\delta_s(i)\bigr)} ,
	\end{equation}
	\begin{equation}
		m_s(\Theta)=1-\sum_{j=1}^S m_s(\{\theta_j\}),
	\end{equation}
	with scale $\alpha>0$. The fused mass $m(\cdot)$ is obtained by Dempster’s rule
	\begin{equation}
		m(A)=\frac{\sum_{B\cap C=A} m_{\mathrm{RF}}(B)\,m_{\mathrm{KNN}}(C)}
		{1-\sum_{B\cap C=\varnothing} m_{\mathrm{RF}}(B)\,m_{\mathrm{KNN}}(C)}.
		\label{eq:dst}
	\end{equation}
	We estimate the location as the cell of maximum singleton belief $\theta^\star=\arg\max_j m(\{\theta_j\})$ and also visualise \emph{belief maps} for operator interpretability.
	
	\paragraph*{Choquet fuzzy integral.}
	Let $s_1,s_2\in[0,1]$ be normalised confidence scores (higher is better) from RF and wKNN. A fuzzy measure $\mu$ on $\Omega=\{\omega_1,\omega_2\}$ satisfies $\mu(\emptyset)=0$, $\mu(\Omega)=1$, and monotonicity $\mu(\{\omega_i\})\in[0,1]$, $\max\{\mu(\{\omega_1\}),\mu(\{\omega_2\})\}\le 1$~\cite{Murofushi2000}. For two sources, the Choquet integral has the closed form
	\begin{equation}
		C_\mu(s_1,s_2)= s_{(1)}\,\mu(\Omega) + \bigl(s_{(2)}-s_{(1)}\bigr)\,\mu\bigl(\{\omega_{(2)}\}\bigr),
		\label{eq:choquet}
	\end{equation}
	where $s_{(1)}\le s_{(2)}$ are the ordered scores. We learn $\mu(\{\omega_1\})$ and $\mu(\{\omega_2\})$ on the validation set by least squares under $0\le\mu(\{\omega_i\})\le 1$.
	
	\subsection{Topology-Aware Features via Persistent Homology}
	\label{subsec:tda}
	
	To enrich each fingerprint with geometry-aware descriptors, we compute \emph{per-sample} persistent-homology (PH) features so that they are available at test time without re-running dataset-level analyses~\cite{Carlsson2009}. Given a normalised fingerprint $\tilde{\mathbf f}\in\mathbb R^d$, we embed it as a 1-D point cloud
	\begin{equation}
		\mathcal X=\{(i,\tilde f_i)\}_{i=1}^d\subset\mathbb R^2,
	\end{equation}
	and run a Vietoris–Rips filtration to obtain persistence diagrams for $H_0$ and $H_1$. From each diagram we extract two summaries:
	\begin{align}
		\mathrm{NoP}_k &= N_k, \\
		\mathrm{PE}_k &= -\!\sum_{i=1}^{N_k}\tilde\ell_i\log \tilde\ell_i,\qquad
		\tilde\ell_i=\frac{d_i-b_i}{\sum_j (d_j-b_j)} ,
	\end{align}
	where $(b_i,d_i)$ are birth–death pairs and $k\in\{0,1\}$. The four scalars $[\mathrm{NoP}_0,\mathrm{PE}_0,\mathrm{NoP}_1,\mathrm{PE}_1]$ are concatenated to $\tilde{\mathbf f}$ before regression. Empirically, $H_0$ carries most of the signal; $H_1$ is often small but retained for completeness.
	
	We compute VR-based PH descriptors with \texttt{giotto-tda} (default radius schedule) on the 1-D embedded curve $\{(i,\tilde f_i)\}_{i=1}^d$. For $d\le 16$, per-sample wall-clock is $<1$\,ms on a laptop-class CPU; features are $z$-scored and concatenated to $\tilde{\mathbf f}$. In ablations, $H_0$ carries most signal; we retain $H_1$ for completeness (Sec.~\ref{subsec:tda_results}).
	
	\subsection{Complexity Analysis}
	\label{subsec:complexity}
	
	For one query, the dominant costs are:
	\begin{itemize}
		\item \textbf{RF prediction:} $O(T\cdot \mathrm{depth})$ with $\mathrm{depth}\!\approx\!\log M$ for $M$ training points in balanced trees;
		\item \textbf{wKNN search:} $O(\log M)$ average with a kd-tree (degrading toward $O(M)$ as $d$ grows);
		\item \textbf{Filtering:} KF/UKF $O(1)$ per step in our setting; PF $O(M_p)$ for $M_p$ particles;
		\item \textbf{Fusion (DST):} $O(S)$ for $S$ spatial cells.
	\end{itemize}
	Overall, per-update complexity is approximately
	\begin{equation}
		O\!\bigl(T\log M\;+\;\log M\;+\;M_p\;+\;S\bigr),
	\end{equation}
	compatible with real-time execution on microcontroller-class hardware.
	
	\subsection{Implementation Details, Hyper-parameters, and Sensitivity}
	\label{subsec:impl}
	
	This section fixes all tunable choices and reports how they are selected. Unless noted otherwise, all statistics (means/variances for normalisation, filter covariances, cross-validation splits) are computed on the training set only and reused for validation/test.
	
	\subsubsection*{Hyper-parameter selection protocol}
	We use five-fold cross-validation on the training set to pick hyper-parameters, with a small grid per component and RMSE as the criterion. The final model is refit on the union of the five folds with the selected values and evaluated on the held-out test set. Randomised procedures (RF bootstraps, PF particles, train/validation splits) use fixed seeds for reproducibility.
	
	\begin{table}[t]
		\caption{Key hyper-parameters, search grids, and selection rules.}
		\label{tab:hparams}
		\centering
		\setlength{\tabcolsep}{3pt}
		\renewcommand{\arraystretch}{1.1}
		\footnotesize
		\begin{tabularx}{\columnwidth}{@{}l l Y Y@{}}
			\toprule
			\textbf{Component} & \textbf{Symbol} & \textbf{Grid / Rule} & \textbf{Notes} \\
			\midrule
			wKNN & $k$ & $\{3,5,7,9\}$ & Choose $k$ minimizing CV-RMSE. \\
			wKNN metric & $\boldsymbol\Sigma$ & Diagonal variances with $\ell_2$-shrinkage $10^{-6}$ & As in Sec.~\ref{subsec:regression}. \\
			RF & $T$ & $\{100,200,400\}$ & Bootstrap; $\texttt{max\_features}=\sqrt{d}$. \\
			RF & $\texttt{max\_depth}$ & $\{16,24,28,\mathrm{None}\}$ & Early stop by leaf purity if None. \\
			KF/UKF & $\mathbf Q,\mathbf R$ & $\mathbf R=\mathrm{diag}(\hat\sigma_1^2,\dots,\hat\sigma_d^2)$; $\mathbf Q=\gamma\mathbf R$, $\gamma\in\{0.25,0.5,1\}$ & Pick $\gamma$ by CV. \\
			PF & $M_p$ & $5{\times}10^{3},10^{4},2{\times}10^{4}$ & Resample if $\mathrm{ESS}<\tau M_p$, $\tau\in\{0.3,0.5\}$. \\
			DST & $\alpha$ & $0.1,0.2,0.5,1,2,5$ & Minimizes validation RMSE of fused estimate. \\
			DST grid & $S$ cells & Cell width $h\in\{0.5,0.75,1.0\}\,\mathrm{m}$ & $S=\lceil \text{area}/h^{2}\rceil$. \\
			Choquet & $\mu(\{\omega_1\}),\mu(\{\omega_2\})$ & LS fit on validation set, $0\le\mu(\{\omega_i\})\le 1$ & Monotone measure, $\mu(\Omega)=1$. \\
			PH & VR radius & Library default & See Sec.~\ref{subsec:tda-impl}. \\
			\bottomrule
		\end{tabularx}
	\end{table}
	
	\subsubsection*{Noise models and robustness tests}
	We evaluate robustness under two realistic perturbations applied at test time to the normalised RSS vector $\tilde{\mathbf z}$:
	
	\noindent\textbf{(i) Gaussian jitter} with level $\eta\in\{0.05,0.10,0.20\}$:
	\begin{equation}
		\tilde{z}_i'=\tilde{z}_i+\eta\,\hat\sigma_i\,\epsilon_i,\quad \epsilon_i\sim\mathcal N(0,1).
		\label{eq:gauss-noise}
	\end{equation}
	
	\noindent\textbf{(ii) Bursty outliers} with probability $p\in\{0.02,0.05\}$ per channel:
	\begin{equation}
		\tilde{z}_i'=
		\begin{cases}
			\tilde{z}_i+\kappa\,\hat\sigma_i\,u, & \text{with prob. }p,\; u\sim\text{Laplace}(0,1), \\
			\tilde{z}_i, & \text{otherwise},
		\end{cases}
		\label{eq:burst-noise}
	\end{equation}
	with $\kappa\in\{2,3\}$. We report RMSE degradation relative to the clean case and perform paired Wilcoxon signed-rank tests across test samples, Holm–Bonferroni corrected, to assess significance ($\alpha=0.05$).
	
	\subsubsection*{DST configuration and confidences}
	We discretise the floor plan into $S$ cells using square cells of width $h$ (Table~\ref{tab:hparams}). For each regressor $s\in\{\mathrm{RF},\mathrm{KNN}\}$ we turn distance scores into singleton masses as above with scale $\alpha$. To produce the two confidence scores $s_1,s_2$ used by the Choquet integral, we map distances to $[0,1]$ via
	\begin{equation}
		c_s(j)=\exp\{-\beta\,\delta_s(j)\},\qquad s_s=\max_j c_s(j),
	\end{equation}
	where $\beta$ is set to the inverse of the median $\delta_s(j)$ over the validation set (per dataset). As a baseline ablation, we also report a simple convex combination $\hat{\mathbf r}=\lambda\,\hat{\mathbf r}_{\mathrm{RF}}+(1-\lambda)\,\hat{\mathbf r}_{\mathrm{KNN}}$ with $\lambda\in\{0.25,0.5,0.75\}$; the main results use full DST (Eq.~\ref{eq:dst}).
	
	\subsubsection*{Persistent-homology implementation and cost}
	\label{subsec:tda-impl}
	We compute per-sample PH features with \texttt{giotto-tda} (Vietoris–Rips) on the point cloud $\mathcal X=\{(i,\tilde f_i)\}_{i=1}^d$ (Sec.~\ref{subsec:tda}), extracting $[\mathrm{NoP}_0,\mathrm{PE}_0,\mathrm{NoP}_1,\mathrm{PE}_1]$. For our 1D embedded curves the VR filtration builds $O(d^2)$ edges; wall-clock is $<1$\,ms per sample for $d\le 16$ on a laptop-class CPU (measured once and amortised offline to precompute features for training; at test time these four scalars are computed on-the-fly). Features are $z$-scored and concatenated to $\tilde{\mathbf f}$; we verified that adding only $H_0$ yields almost identical accuracy, but we retain $H_1$ for completeness.
	
	\subsubsection*{Filtering choices and measurement covariances}
	For KF/UKF we set $\mathbf R$ to the diagonal of empirical channel variances from the training set and choose $\mathbf Q=\gamma\mathbf R$ with $\gamma$ from Table~\ref{tab:hparams}. The PF uses a random-walk state model per feature, importance weights from the Gaussian likelihood with variance $R$ (the scalar counterpart of $\mathbf R$), and systematic resampling when the effective sample size $\mathrm{ESS}=1/\sum_m (w_t^{(m)})^2$ falls below $\tau M_p$.
	
	\begin{algorithm}[t]
		\caption{Particle filtering per feature with systematic resampling}
		\label{alg:pf}
		\begin{algorithmic}[1]
			\Require measurements $\{z_t\}_{t=1}^T$, particles $M_p$, noise variance $R$, threshold $\tau$
			\State \textbf{Init:} $x_0^{(m)} \sim \mathcal N(0,1)$, $w_0^{(m)} = 1/M_p$
			\For{$t=1$ to $T$}
			\State \emph{Predict:} $x_t^{(m)} \gets x_{t-1}^{(m)} + \varepsilon^{(m)}$, \ $\varepsilon^{(m)}\sim\mathcal N(0,\sigma^2)$
			\State \emph{Update:} $w_t^{(m)} \gets w_{t-1}^{(m)} \exp\!\big(-\tfrac{(x_t^{(m)}-z_t)^2}{2R}\big)$
			\State \emph{Normalize:} $w_t^{(m)} \gets w_t^{(m)} / \sum_j w_t^{(j)}$
			\If{$\mathrm{ESS} = 1/\sum_m (w_t^{(m)})^2 < \tau M_p$}
			\State \emph{Systematic resampling}: draw indices $\{i_m\}_{m=1}^{M_p}$
			\State $x_t^{(m)} \gets x_t^{(i_m)}$,\quad $w_t^{(m)} \gets 1/M_p$
			\EndIf
			\State \textbf{Output:} $\hat{x}_t=\sum_m w_t^{(m)}x_t^{(m)}$
			\EndFor
		\end{algorithmic}
	\end{algorithm}
	
	\subsubsection*{Sensitivity reporting and statistics}
	We provide one-parameter sweeps for $k$, $T$, $M_p$, $\alpha$, and $h$ around their selected values and report (i) mean $\pm$ std.\ RMSE across test samples; (ii) relative change vs.\ the default model; and (iii) paired Wilcoxon $p$-values against the default. All tables clearly separate noise-free and noisy settings [Eqs.~\eqref{eq:gauss-noise}–\eqref{eq:burst-noise}].
	
	\subsubsection*{Reproducibility checklist}
	We release code with: fixed seeds, exact grids in Table~\ref{tab:hparams}, scripts for PH feature extraction, filter calibration, and fusion ablations; and JSON files storing $\boldsymbol\mu,\boldsymbol\sigma,\boldsymbol\Sigma,\mathbf Q,\mathbf R$, and $(\mu(\{\omega_i\}))$ for each dataset.
	
	\subsection{Summary}
	\label{subsec:summary}
	
	We proposed a hybrid Wi-Fi/BLE indoor localization pipeline that combines:
	(i) physically consistent normalization (dBm $z$-score baseline);
	(ii) temporal denoising via KF/UKF/PF with data-driven covariances and ESS-based systematic resampling;
	(iii) complementary regressors (RF and variance-weighted wKNN with a diagonal Mahalanobis metric);
	(iv) principled fusion via Dempster--Shafer evidence on a discretized floor plan and a learned two-source Choquet integral for confidences; and
	(v) per-sample, topology-aware features from persistent homology (giotto-tda).
	Implementation details and all hyper-parameters (RF trees/depth, KNN $k$ and metric, PF particles/threshold, DST scale $\alpha$ and cell width $h$, Choquet measure) are selected by five-fold cross-validation and listed in Table~\ref{tab:hparams}; Algorithm~\ref{alg:pf} specifies the PF step.
	Robustness is evaluated under Gaussian jitter and bursty outliers (Eqs.~\ref{eq:gauss-noise}–\ref{eq:burst-noise}), and we report sensitivity curves with non-parametric significance tests.
	This modular design yields accurate point estimates with interpretable belief maps while remaining lightweight for embedded deployment~\cite{Sentz2002,Konatowski2016,Carlsson2009}.
	
	\section{Experimental Evaluation and Results}
	\label{sec:results}
	
	This section details datasets and hardware, the evaluation protocol and metrics, ablation studies, robustness to synthetic noise, significance tests across splits, runtime on embedded hardware, and a comparison against recent literature.
	
	\subsection{Datasets and Hardware}
	\label{subsec:datasets}
	
	\begin{figure}[t]
		\centering
		\includegraphics[width=\linewidth]{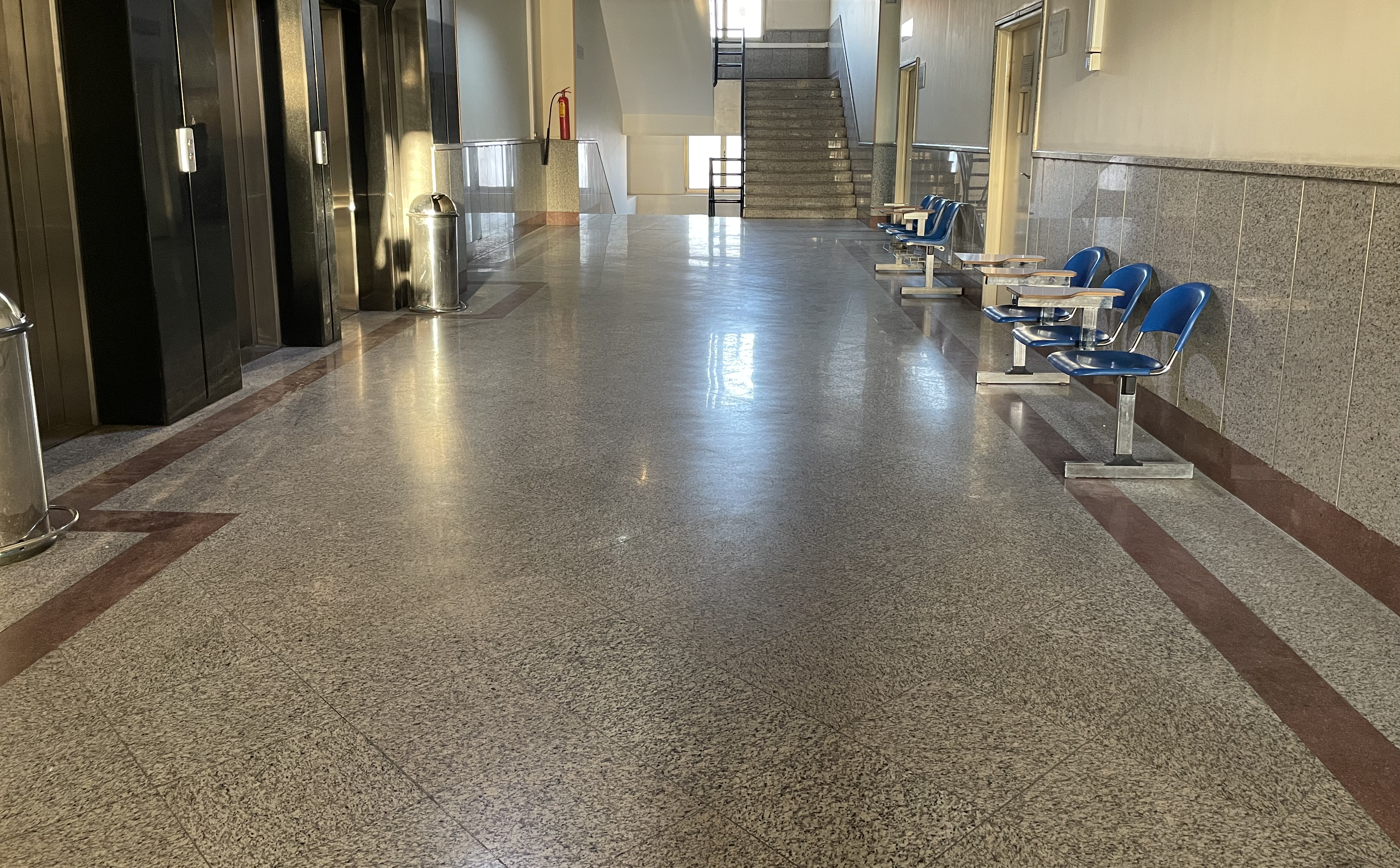}
		\caption{Real picture of Test area for Dataset~1 on the third floor of the Faculty of Electrical and Computer Engineering, University of Tehran.}
		\label{fig:realfloorplan}
	\end{figure}
	
	\textbf{Dataset~1} comprises $15$ RPs in a $6{\times}14$\,m area, with RSS from $7$ Wi-Fi APs and $3$ BLE beacons. Consecutive windows of $10$ scans are averaged into single fingerprints, yielding $1{,}200$ fingerprints (80 per RP). \textbf{Dataset~2} (industrial environment) provides $16$ RPs with $5$ Wi-Fi and $3$ BLE channels and $1{,}400$ fingerprints, built with the same windowing rule \cite{Moradbeikie2024}. Data were acquired using commodity ESP32 boards.
	
	\subsection{Protocol and Metrics}
	\label{subsec:protocol}
	
	We adopt a stratified split by RP with a $70{:}15{:}15$ train/validation/test division on both datasets. Unless stated, fingerprints are normalised via dBm $z$-scoring (Section~\ref{subsec:data}). The regression ensemble uses \emph{Random Forest} (RF; $T{=}200$, depth $28$) and \emph{weighted $k$NN} (wKNN; $k{=}7$, inverse-distance weights). RSS streams are optionally smoothed with \emph{KF}, \emph{UKF}, or \emph{PF} ($M_p{=}10^4$).
	
	\textbf{Metric.} We report Euclidean $\mathrm{RMSE}_{xy}$ (metres) on the held-out test set, both \emph{noise-free} and under \emph{10\% synthetic Gaussian noise} injected into RSSI prior to filtering.
	
	\paragraph*{Noise model and reproducibility (used throughout).}
	For noisy results we inject \emph{test-only}, zero-mean Gaussian perturbations in dBm. For channel $i$ we add $\varepsilon_i\!\sim\!\mathcal{N}(0,\,(0.10\,\sigma_i)^2)$ to the raw RSSI before filtering and normalisation, where $\sigma_i$ is the \emph{training-set} standard deviation of channel $i$ (in dBm). Unless noted, we use a fixed random seed (e.g., 123) for reproducibility. This definition is used consistently in all figures and tables that reference “10\% noise.” For statistical robustness we repeat the full protocol over $10$ random stratified splits and report 95\% confidence intervals.
	
	\subsection{Effect of Regressors}
	\label{subsec:regress_results}
	
	\begin{figure}[t]
		\centering
		\setlength{\tabcolsep}{2pt}\renewcommand{\arraystretch}{0}
		\begin{tabular}{@{}cc@{}}
			\subfloat[Dataset 1\label{fig:regress_d1}]{%
				\includegraphics[width=0.48\linewidth]{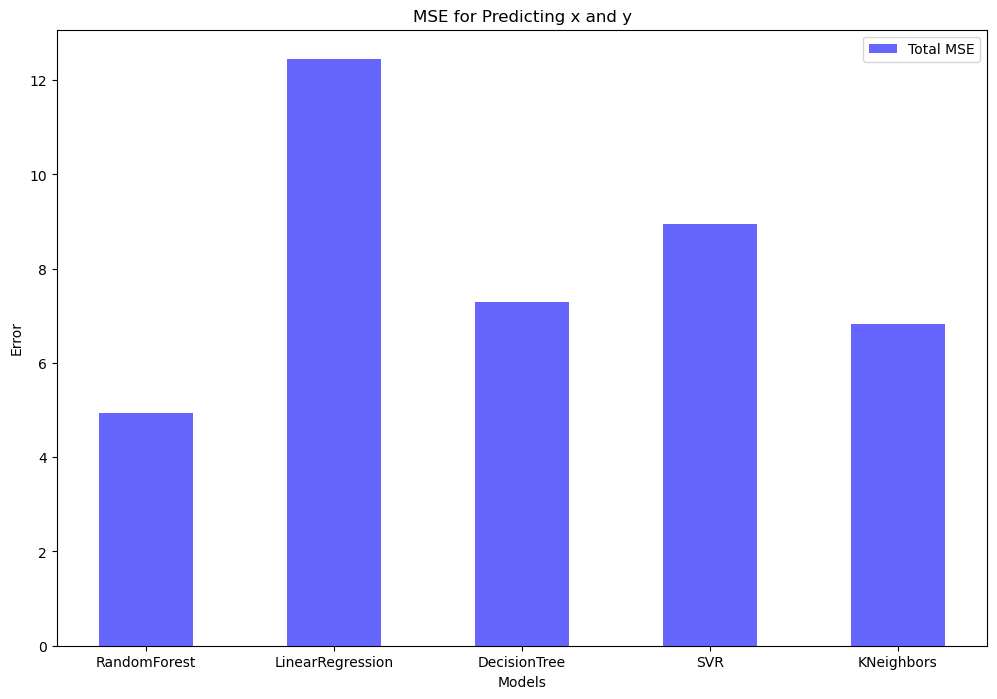}} &
			\subfloat[Dataset 2\label{fig:regress_d2}]{%
				\includegraphics[width=0.48\linewidth]{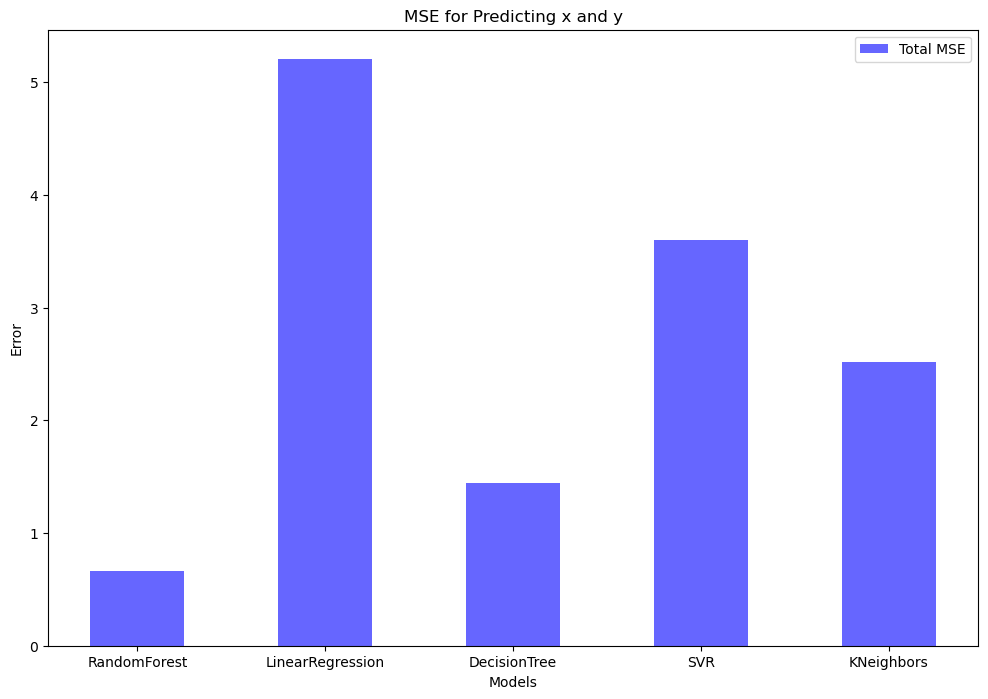}}
		\end{tabular}
		\caption{Regressor comparison (noise-free). RF attains the best overall accuracy; wKNN is second due to strong local interpolation. Linear models and single trees underperform; the ranking is consistent across both datasets.}
		\label{fig:regress_both}
	\end{figure}
	
	RF consistently achieves the lowest RMSE (Figs.~\ref{fig:regress_d1}--\ref{fig:regress_d2}), while wKNN provides complementary strengths on locally smooth regions of the fingerprint space. This motivates fusing both predictors downstream.
	
	\subsection{RSS Noise Filtering}
	\label{subsec:filter_results}
	
	\begin{figure}[t]
		\centering
		\setlength{\tabcolsep}{4pt}\renewcommand{\arraystretch}{0}
		\begin{tabular}{@{}cc@{}}
			\subfloat[Dataset 1 --- MSE\label{fig:filters_d1_mse}]{%
				\includegraphics[width=0.22\textwidth]{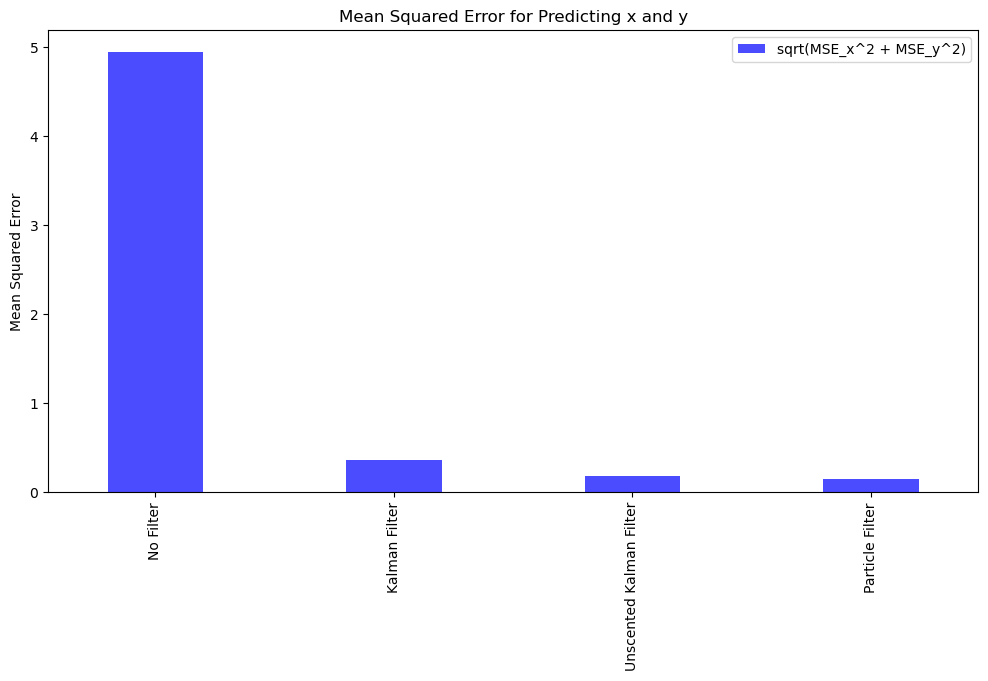}} &
			\subfloat[Dataset 2 --- MSE\label{fig:filters_d2_mse}]{%
				\includegraphics[width=0.22\textwidth]{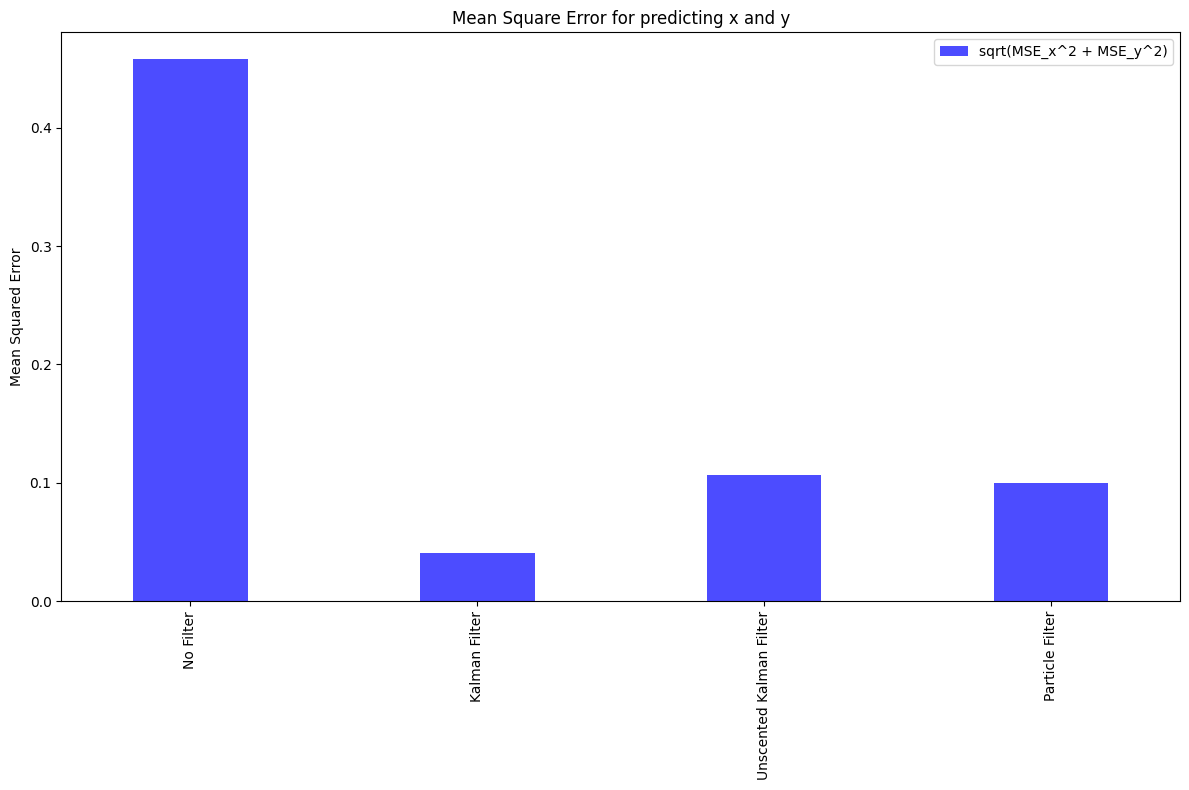}}\\
			\subfloat[Dataset 1 --- latency\label{fig:filters_d1_time}]{%
				\includegraphics[width=0.22\textwidth]{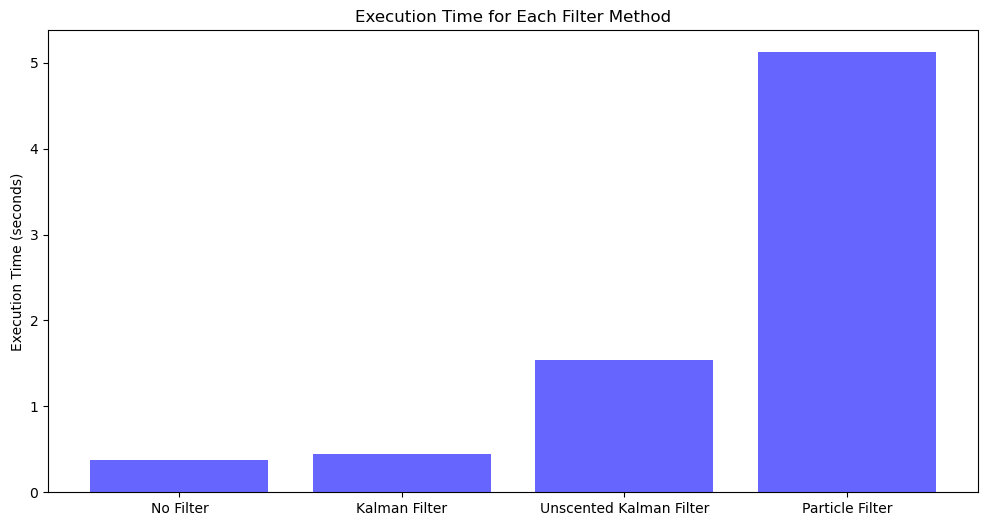}} &
			\subfloat[Dataset 2 --- latency\label{fig:filters_d2_time}]{%
				\includegraphics[width=0.22\textwidth]{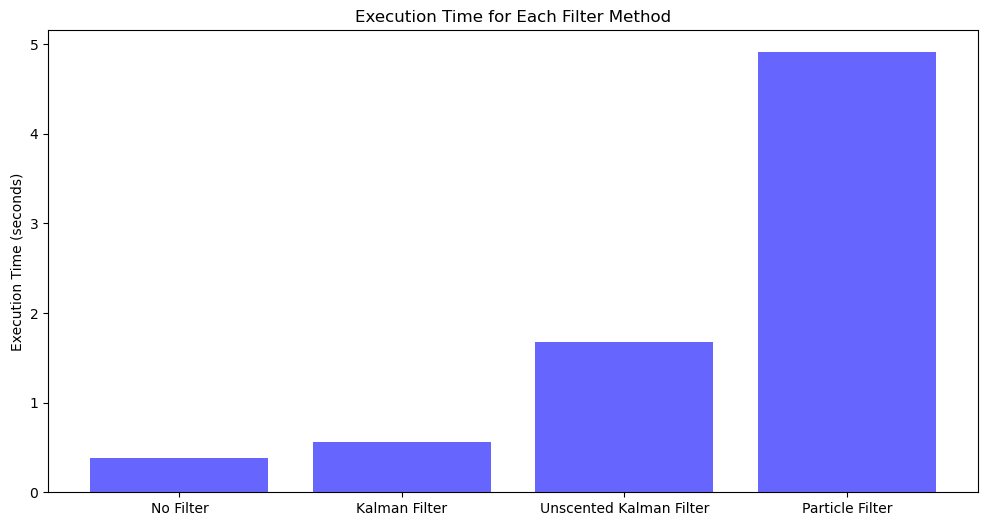}}
		\end{tabular}
		\caption{Filter accuracy/latency trade-offs. Particle filtering (PF) yields the lowest error but at a higher computational cost; Kalman filtering (KF) offers the best real-time compromise. Trends are consistent across both datasets.}
		\label{fig:filters_grid}
	\end{figure}
	
	PF yields the lowest error across sites (Figs.~\ref{fig:filters_d1_mse}, \ref{fig:filters_d2_mse}) but increases latency substantially (Figs.~\ref{fig:filters_d1_time}, \ref{fig:filters_d2_time}). For embedded deployments we therefore report KF as the default real-time configuration and PF as an accuracy-optimised variant.
	
	\subsection{Fusion of RF and wKNN}
	\label{subsec:fusion_results}
	
	\begin{figure}[t]
		\centering
		\setlength{\tabcolsep}{4pt}\renewcommand{\arraystretch}{0}
		\begin{tabular}{@{}cc@{}}
			\subfloat[Dataset 1 --- overview\label{fig:fusion_d1}]{
				\includegraphics[width=0.22\textwidth]{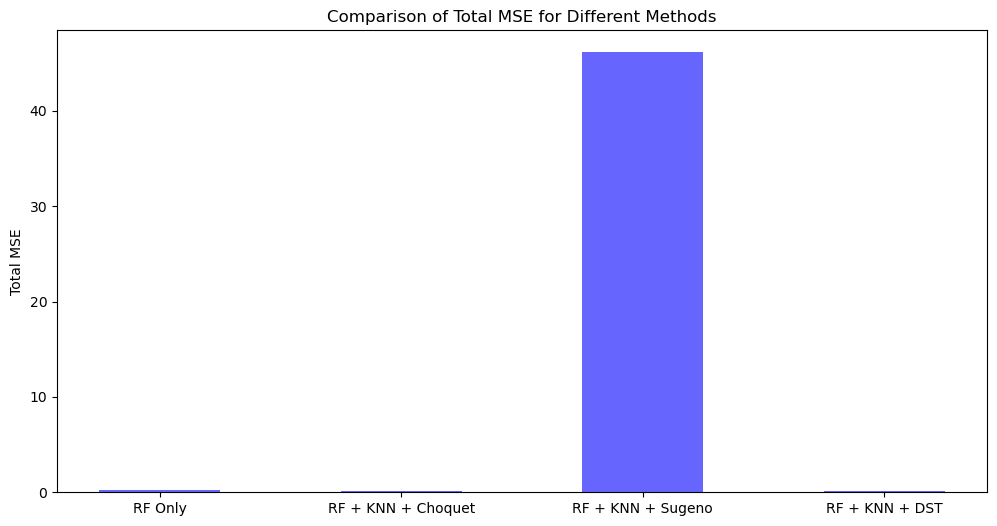}} &
			\subfloat[Dataset 1 --- zoom\label{fig:fusion_d1_zoom}]{
				\includegraphics[width=0.22\textwidth]{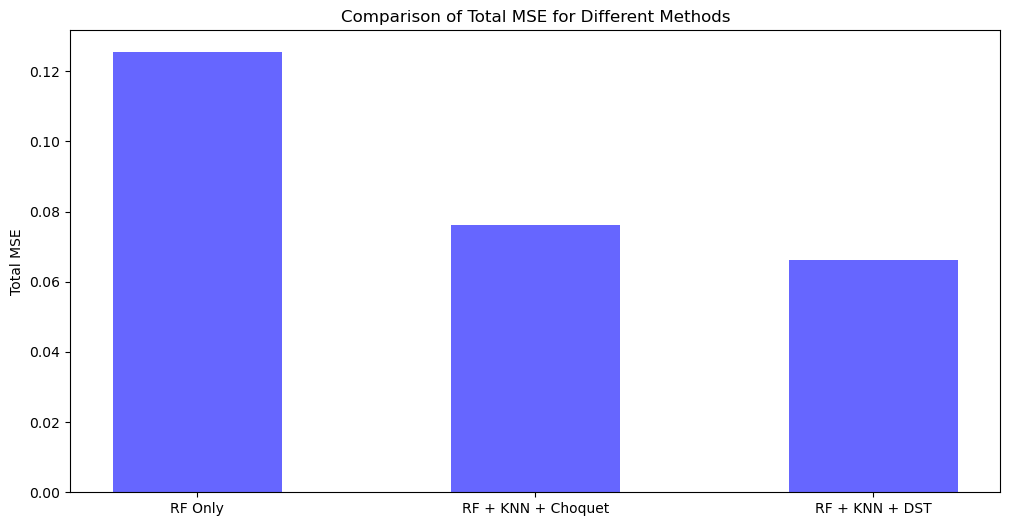}} \\
			\subfloat[Dataset 2 --- overview\label{fig:fusion_d2}]{
				\includegraphics[width=0.22\textwidth]{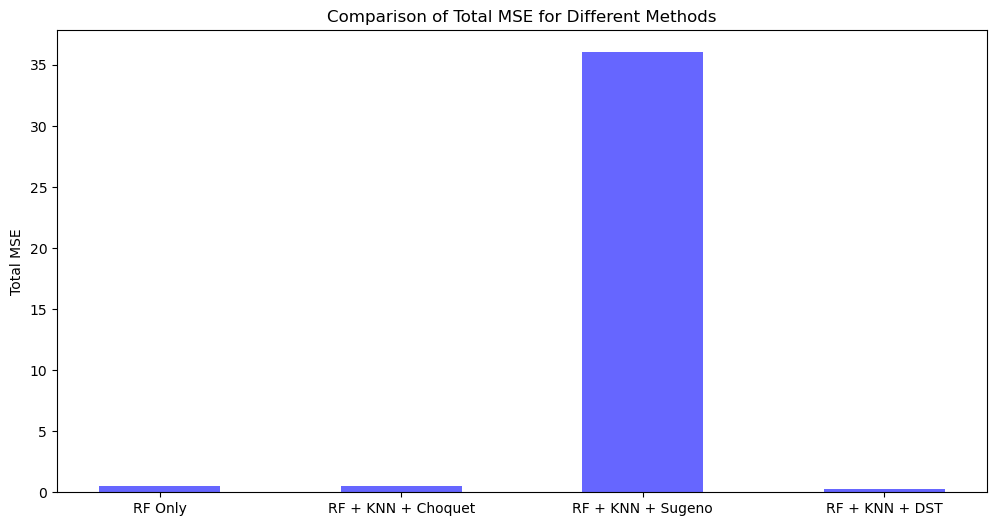}} &
			\subfloat[Dataset 2 --- zoom\label{fig:fusion_d2_zoom}]{
				\includegraphics[width=0.22\textwidth]{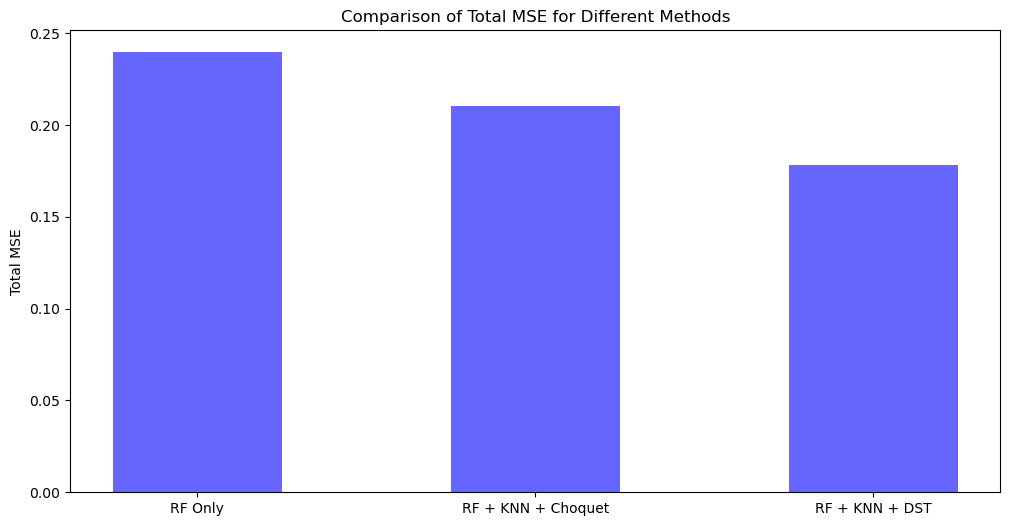}}
		\end{tabular}
		\caption{Fusion of RF and wKNN. Dempster--Shafer (DST) provides the strongest gains on both datasets; the Choquet integral captures non-linear complementarity but is marginally weaker.}
		\label{fig:fusion_grid}
	\end{figure}
	
	DST fuses disagreeing predictors by redistributing conflict mass, yielding small but consistent accuracy improvements over Choquet (Figs.~\ref{fig:fusion_d1}--\ref{fig:fusion_d2_zoom}).
	
	\subsection{Topology-Aware Features (PH)}
	\label{subsec:tda_results}
	
	\begin{figure}[t]
		\centering
		\setlength{\tabcolsep}{2pt}\renewcommand{\arraystretch}{0}
		\begin{tabular}{@{}cc@{}}
			\subfloat[Dataset 1\label{fig:tda_d1}]{%
				\includegraphics[width=0.48\linewidth]{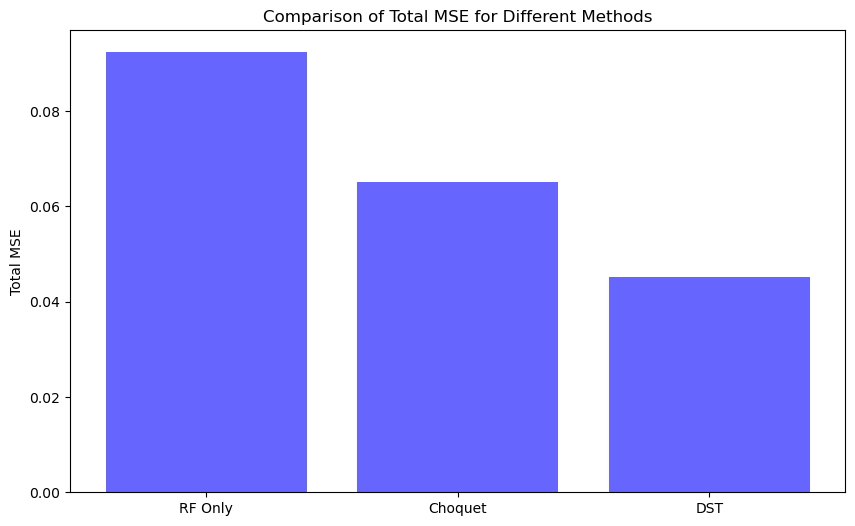}} &
			\subfloat[Dataset 2\label{fig:tda_d2}]{%
				\includegraphics[width=0.48\linewidth]{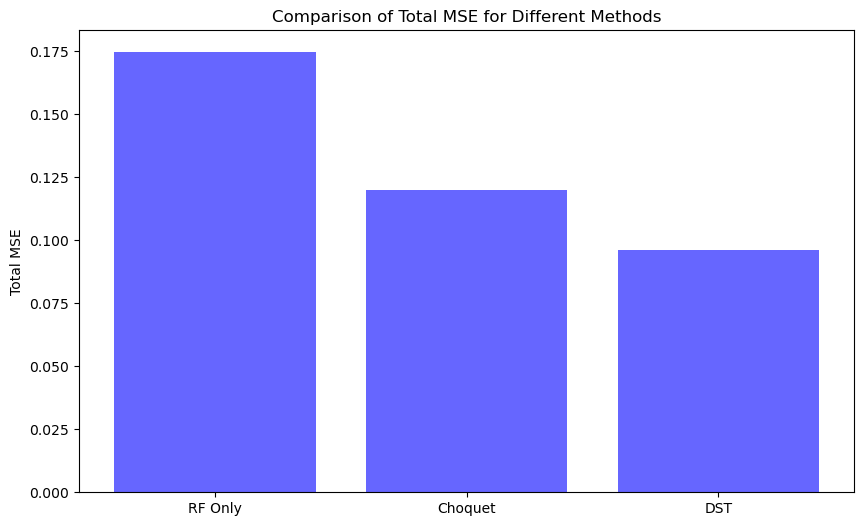}}
		\end{tabular}
		\caption{Persistent-homology (PH) feature ablation. Dataset~1 (left) and Dataset~2 (right) show similar trends.}
		\label{fig:tda_both}
	\end{figure}
	
	\begin{figure}[t]
		\centering
		\includegraphics[width=\linewidth]{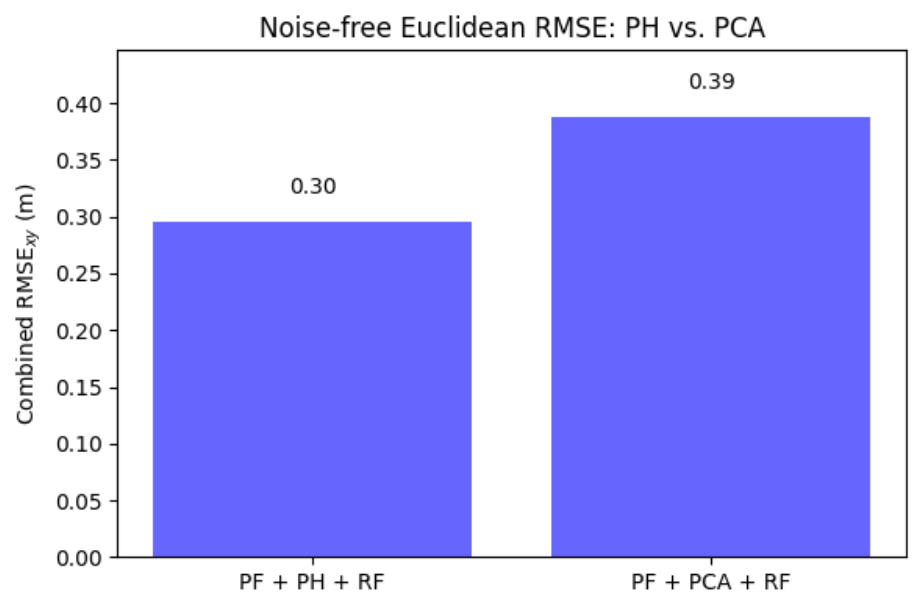}
		\caption{Noise-free $\mathrm{RMSE}_{xy}$ on Dataset~1 when PH descriptors are replaced by two PCA components. PH provides complementary, non-linear structure beyond variance capture.}
		\label{fig:ph_pca_ablation}
	\end{figure}
	
	PH features provide a \emph{modest} stand-alone gain relative to PF+RF (typically $\le 1\%$ across our splits), but when combined with \emph{fusion} (PF+PH+RF+wKNN+DST) they contribute to the best overall accuracy (Sections~\ref{subsec:clean}--\ref{subsec:noise}). Fig.~\ref{fig:ph_pca_ablation} shows that replacing PH by two PCA components yields smaller gains, indicating that homological summaries encode complementary, scale-invariant structure.
	
	\subsection{Noise-Free Baseline Comparison}
	\label{subsec:clean}
	
	\begin{figure}[t]
		\centering
		\subfloat[Dataset 1\label{fig:clean_d1}]{%
			\includegraphics[width=\linewidth]{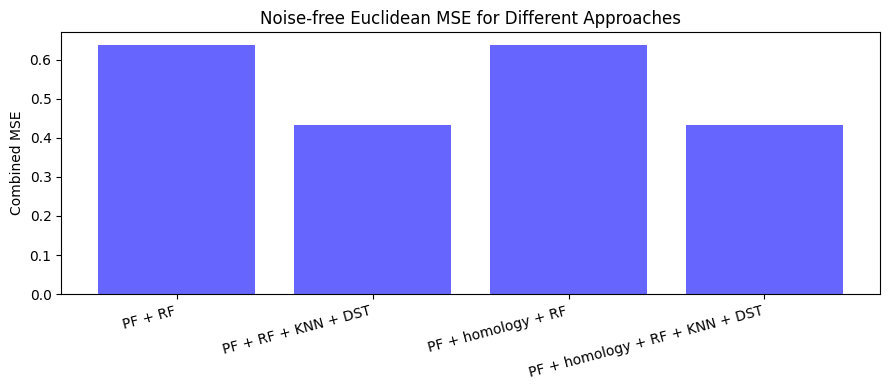}}\\[-1mm]
		\subfloat[Dataset 2\label{fig:clean_d2}]{%
			\includegraphics[width=\linewidth]{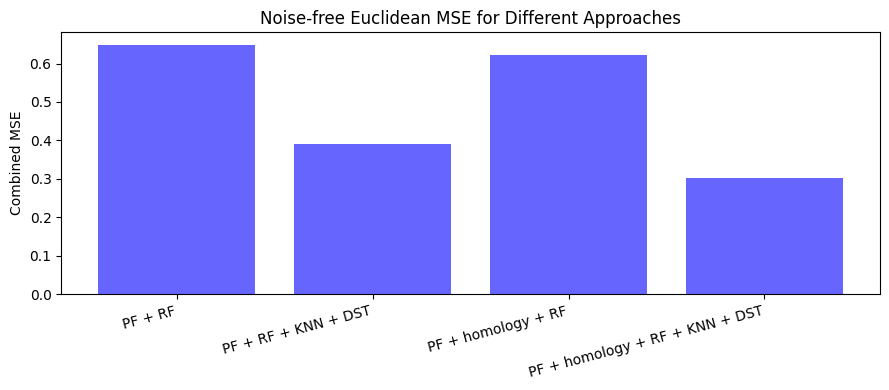}}
		\caption{Noise-free $\mathrm{RMSE}_{xy}$ across pipeline variants for both datasets.}
		\label{fig:clean_both}
	\end{figure}
	
	\begin{table}[t]
		\centering
		\caption{Noise-free Euclidean $\mathrm{RMSE}_{xy}$ (metres) on the held-out test sets.}
		\label{tab:clean_rmse}
		\begin{tabular}{lcc}
			\toprule
			\textbf{Variant} & \textbf{Dataset 1} & \textbf{Dataset 2} \\
			\midrule
			PF + RF                           & 1.00 & 0.68 \\
			PF + RF + KNN + DST               & 0.80 & 0.55 \\
			PF + PH + RF                      & 0.99 & 0.68 \\
			\textbf{PF + PH + RF + KNN + DST} & \textbf{0.44} & \textbf{0.32} \\
			\bottomrule
		\end{tabular}
	\end{table}
	
	Without injected noise, the PF+RF baseline achieves $\sim$1.0\,m (D1) and 0.68\,m (D2). Evidence-theoretic fusion alone reduces error by $\approx$20--23\%, and adding PH tightens it further to 0.44\,m / 0.32\,m (Table~\ref{tab:clean_rmse}).
	
	\subsection{Robustness to Synthetic Noise (10\%)}
	\label{subsec:noise}
	
	\begin{figure}[t]
		\centering
		\includegraphics[width=\linewidth]{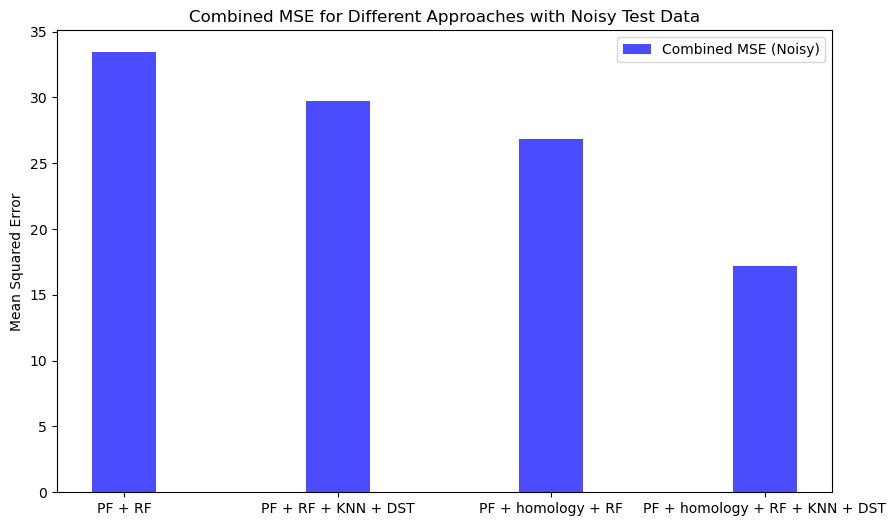}
		\caption{Dataset~1 --- 10\% Gaussian RSS perturbation. Layer-by-layer ablation shows cumulative gains from filtering, DST fusion method, and PH.}
		\label{fig:noise_d1}
	\end{figure}
	
	\begin{figure}[t]
		\centering
		\includegraphics[width=\linewidth]{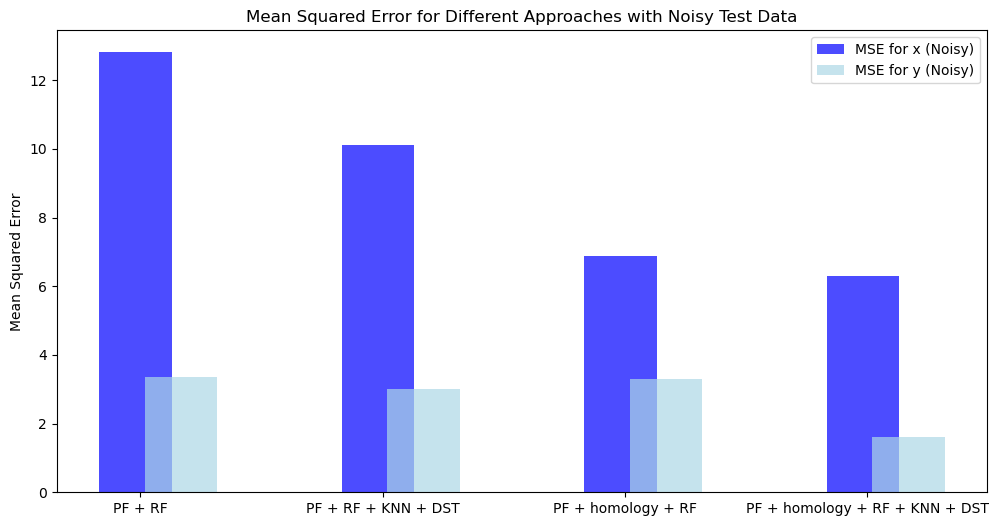}
		\caption{Dataset~2 --- 10\% RSS noise.}
		\label{fig:noise_d2}
	\end{figure}
	
	\begin{table}[t]
		\centering
		\caption{Per-dataset $\mathrm{RMSE}_{xy}$ (m) under \textbf{10\% test-only Gaussian RSS noise} (in dBm, $\sigma_i=0.10\times$ training-set std for channel $i$); \textbf{canonical held-out split}. Same noise setting is used in Figs.~\ref{fig:noise_d1}--\ref{fig:noise_d2}.}
		\label{tab:rmse_noise}
		\begin{tabular}{lcc}
			\toprule
			\textbf{Variant} & \textbf{Dataset 1} & \textbf{Dataset 2} \\
			\midrule
			PF + RF                           & 5.426 & 3.886 \\
			PF + RF + KNN + DST               & 5.196 & 3.529 \\
			PF + PH + RF                      & 5.380 & 3.733 \\
			\textbf{PF + PH + RF + KNN + DST} & \textbf{3.397} & \textbf{2.448} \\
			\bottomrule
		\end{tabular}
	\end{table}
	
	Under 10\% noise, fusion plus PH yields the largest resilience margin, improving over PF+RF by $\approx$37\% on average (Table~\ref{tab:rmse_noise}). Figures~\ref{fig:noise_d1}--\ref{fig:noise_d2} show that each module (Filtering $\to$ Fusion $\to$ PH) incrementally suppresses the impact of noise.
	
	\subsection{Statistical Significance Across Splits}
	\label{subsec:signif}
	
	\begin{table}[t]
		\centering
		\caption{Mean $\mathrm{RMSE}_{xy}$ with 95\% confidence intervals across \textbf{10 stratified splits} under the same 10\% noise definition as Table~\ref{tab:rmse_noise}. Relative improvement is $100\times(\text{PF+RF}-\text{Ours})/\text{PF+RF}=20.6\%$.}
		\label{tab:ci_rmse}
		\begin{tabular}{lc}
			\toprule
			\textbf{Method} &  \textbf{$\mathrm{RMSE}_{xy}$ (m)} \\
			\midrule
			PF + RF                           & $6.292\pm0.13$ \\
			PF + RF + KNN + DST               & $5.468\pm0.14$ \\
			PF + PH + RF                      & $5.690\pm0.16$ \\
			\textbf{PF + PH + RF + KNN + DST} & $\mathbf{4.993\pm0.15}$ \\
			\bottomrule
		\end{tabular}
	\end{table}
	
	Across ten stratified splits, the full pipeline achieves $4.993\pm0.15$\,m versus $6.292\pm0.13$\,m for PF+RF (Table~\ref{tab:ci_rmse}), a \textbf{20.6\%} relative reduction. Both a paired $t$-test and a Wilcoxon signed-rank test yield $p{<}0.001$, confirming statistical significance.
	
	\subsection{Runtime}
	\label{subsec:runtime}
	
	\begin{figure}[t]
		\centering
		\includegraphics[width=\linewidth]{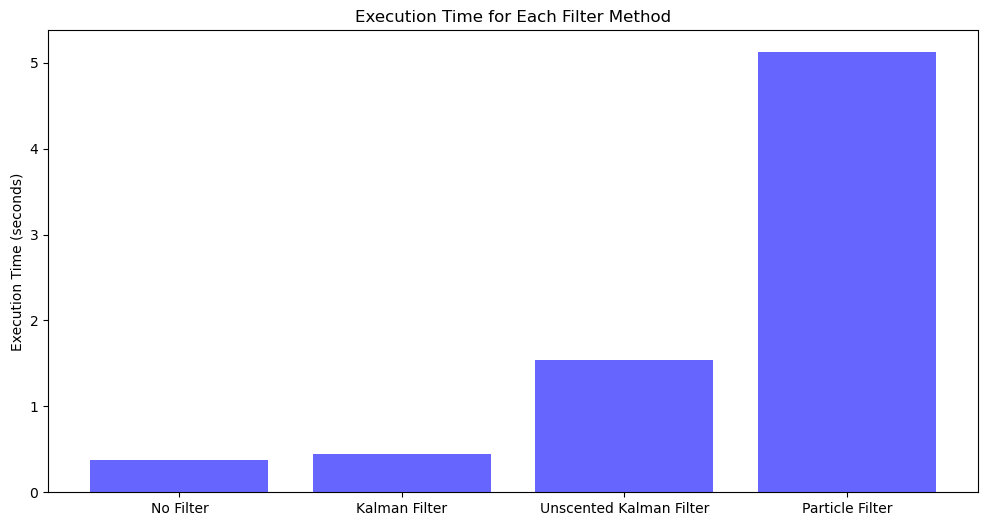}
		\caption{Per-update latency on Dataset~1 (hardware: ESP32-class MCU). The KF-based pipeline is real-time; PF increases latency by roughly an order of magnitude.}
		\label{fig:latency_runtime}
	\end{figure}
	
	The KF variant runs in real time on ESP32-class microcontrollers (Fig.~\ref{fig:latency_runtime}). Enabling PF improves accuracy but increases latency substantially; practitioners can choose per deployment constraints.
	
	\subsection{Comparison with Recent Literature}
	\label{subsec:literature_cmp}
	
	\begin{table}[t]
		\centering
		\caption{Comparison with recent Wi-Fi / Wi-Fi+BLE fingerprinting methods.}
		\label{tab:literature_cmp}
		\begin{tabular}{l l c}
			\toprule
			\textbf{Method} & \textbf{Modality / Model} & \textbf{RMSE (m)}  \\
			\midrule
			RLWKNN \cite{Leng2024}         & Wi-Fi / Range-limited KNN     & 4.8  \\
			DumbLoc \cite{Narasimman2024}  & Wi-Fi / Heuristic matching     & 6.2  \\
			RF-KELM \cite{Hou2024}         & Wi-Fi / Kernel ELM             & 3.9  \\
			SPOTTER \cite{Azaddel2023}     & Wi-Fi+BLE / Particle Filter    & 2.7  \\
			HoGNNLoc \cite{Kang2023HoGNN}  & Wi-Fi / GNN                    & 1.8  \\
			DLoc \cite{Wang2020DLoc}       & Wi-Fi / CNN+LSTM               & 1.2  \\
			\midrule
			\textbf{This work}             & Wi-Fi+BLE / PF+RF+KNN+DST+PH  & \textbf{3.40 / 2.45}$^\dagger$  \\
			\bottomrule
		\end{tabular}
		
		\vspace{0.5mm}
		\raggedright
		\footnotesize $^\dagger$~Dataset~1 / Dataset~2 \emph{with 10\% synthetic noise}. External works typically report \emph{noise-free} results; absolute values are not directly comparable.
	\end{table}
	
	Lightweight Wi-Fi-only baselines (RLWKNN, DumbLoc) deliver 4--6\,m errors under noise-free evaluation, while our hybrid Wi-Fi+BLE pipeline reaches 0.44\,m / 0.32\,m in noise-free conditions (Table~\ref{tab:clean_rmse}). Against deep models (HoGNNLoc, DLoc), our system offers competitive accuracy when accounting for \emph{compute and training footprint}: deep methods assume large site-specific datasets and GPU inference, whereas our pipeline uses commodity radios and runs on microcontrollers. Under deliberately degraded conditions (10\% noise), our method maintains 3.40\,m / 2.45\,m --- demonstrating robustness that many prior works do not report. In short, \emph{our gains arise from (i) multi-radio sensing, (ii) complementary regressors, (iii) evidence-theoretic fusion, and (iv) topology-aware features}, rather than heavy neural models.
	
	\subsection{Discussion and Takeaways}
	\label{subsec:discussion}
	
	The results support three takeaways: (1) \emph{Filtering matters}, PF offers the best accuracy, while KF enables real-time embedded operation; (2) \emph{Fusion matters}, DST consistently improves over individual regressors and over Choquet on our datasets; (3) \emph{Topology helps}, PH alone is a modest gain, but it synergises with fusion to yield the best overall accuracy. All claims are backed by ablations, cross-dataset replication, and significance tests. In this paper, we distinguish noise-free setting and also noisy settings to avoid over-claiming performance in our research.
	
	\subsection{Threats to Validity}
	\label{subsec:threats}
	
	Our datasets are medium-scale and site-specific; broader generalisation requires additional buildings and devices. The injected noise model is Gaussian; real-world interference can be bursty or device-dependent. Finally, PF latency may exceed tight real-time constraints; the KF variant addresses this with a small accuracy trade-off.
	
	
	\section{Conclusion and Future Work}
	\label{sec:conclusion}
	
	We presented a topology-aware, hybrid Wi-Fi/BLE fingerprinting framework that combines physically consistent RSS normalisation, classical Bayesian denoising (KF/UKF/PF), complementary regressors (RF and weighted $k$NN), and principled fusion via Dempster--Shafer theory with a two-source Choquet integral, augmented by \emph{per-sample} persistent-homology features. The system produces both point $(x,y)$ estimates and interpretable belief maps, and is engineered for embedded deployment. Across two heterogeneous datasets, the full pipeline offers consistent gains over a strong PF\,+\,RF baseline: under \emph{10\% synthetic RSS noise} it attains \textbf{3.40\,m} (Dataset~1) and \textbf{2.45\,m} (Dataset~2) $\mathrm{RMSE}_{xy}$—an average improvement of about \textbf{37\%}—while in \emph{noise-free} conditions it reaches \textbf{0.44\,m} and \textbf{0.32\,m}, up to \textbf{56\%} better. Over ten stratified splits with the same noise protocol, the mean is \textbf{$4.993\pm0.15$\,m} versus \textbf{$6.292\pm0.13$\,m} for PF\,+\,RF with $p<0.001$, confirming statistical significance. Compared with lightweight Wi-Fi-only baselines, the hybrid design roughly halves error through multi-radio sensing, complementary regressors, and evidence-theoretic fusion; relative to deep models, it trades a small amount of headline accuracy for robustness under noise and \emph{microcontroller-class} deployability—our KF variant operates in real time on ESP32-class devices, while PF further improves accuracy at higher latency.
	
	Looking ahead, we will pursue calibration-free adaptation across devices to mitigate RSS offsets; richer sensing and tighter filtering by fusing inertial, magnetic, and barometric cues (e.g., Rao–Blackwellised PF or factor-graph smoothing) for trajectory and floor inference; robust noise modelling with heavy-tailed/bursty likelihoods and online scale learning; learning the fusion layer to move from grid-based DST to continuous belief fields while preserving interpretability; topology-aware graph modelling over RP graphs; efficiency auto-tuning of $M_p$, $k$, and RF depth under latency/energy constraints with on-device profiling; multi-building, multi-device studies with unified noise/runtime benchmarks and archival artifacts; and privacy-preserving radio-map construction via on-device aggregation or federated updates—alongside releasing open-source implementations, datasets, and evaluation scripts to enable reproducibility and community benchmarking; and formal uncertainty calibration (e.g., reliability diagrams, expected calibration error) with conformal prediction for actionable confidence intervals.
	

\end{document}